\documentclass{article}
\usepackage{spconf}
\usepackage{amsmath}
\usepackage{graphicx}
\usepackage{booktabs}
\usepackage{makecell}
\usepackage[headsep=1cm,headheight=0cm,left=18mm,right=18mm,bottom=18mm]{geometry} 
\usepackage{fancyhdr}
\usepackage{url}
\usepackage[utf8]{inputenc}

\title{Learning-Based Conditional Image Coder Using Color Separation}

\name{\normalsize Panqi Jia$^{\dagger \ddag}$, Ahmet Burakhan Koyuncu$^{\star \ddag}$, Georgii Gaikov$^{\ddag}$, Alexander Karabutov$^{\ddag}$, Elena Alshina$^{\ddag}$, André Kaup$^{\dagger}$}
\address{\normalsize$^{\dagger}$ Multimedia Communication and Signal Processing, Friedrich-Alexander-University, Germany\\\normalsize$^{\star}$ Chair of Media Technology, Technical University of Munich, Germany\\\normalsize$^{\ddag}$ Huawei Technologies}

\begin{document}
\maketitle
\begin{abstract}
Recently, image compression codecs based on Neural Networks (NN) outperformed the state-of-art classic ones such as BPG, an image format based on HEVC intra. However, the typical NN codec has high complexity, and it has limited options for parallel data processing. In this work, we propose a conditional separation principle that aims to improve parallelization and lower the computational requirements of an NN codec. We present a Conditional Color Separation (CCS) codec which follows this principle. The color components of an image are split into primary and non-primary ones. The processing of each component is done separately, by jointly trained networks. Our approach allows parallel processing of each component, flexibility to select different channel numbers, and an overall complexity reduction. The CCS codec uses over 40\% less memory, has 2x faster encoding and 22\% faster decoding speed, with only 4\% BD-rate loss in RGB PSNR compared to our baseline model over BPG.

\end{abstract}
\pagestyle{fancy}
\fancyhead[ch]{}
\fancyhead{}
\fancypagestyle{FirstPage}{
\lhead{\copyright  2022 IEEE. Personal use of this material is permitted. Permission from IEEE must be obtained for all other uses, in any current or future media, including reprinting/republishing this material for advertising or promotional purposes, creating new collective works, for resale or redistribution to servers or lists, or reuse of any copyrighted component of this work in other works.} 
}
\renewcommand{\headrulewidth}{0pt}
\thispagestyle{FirstPage}
\section{Introduction}
\noindent The aim of image compression is to achieve a good trade-off between a small size of the bitstream and a high quality of the reconstruction. Classic image compression codecs, such as  JPEG \cite{125072}, JPEG2000 \cite{952804}, HEVC \cite{6316136}, BPG \cite{bellard2015bpg}, which is based on HEVC intra, and VVC \cite{9301847} use a combination of fixed block-wise transformation, intra prediction, quantization, arithmetic coders and de-blocking filters to reach this goal.

Neural network-based image and video compression is a more recent research trend. Typically, an NN-based video codec would contain a dedicated image compression network. Three fundamental NN-based image compression concepts were proposed by Balle et al. - \textit{autoencoder} \cite{Ball2017EndtoendOI}, \textit{hyperprior} \cite{Ball2018VariationalIC} and \textit{hyperprior with context model} \cite{Minnen2018JointAA}. In those works, the parts of the codec are jointly trained. Building on top of Balle's work, Zhou et al. \cite{Zhou2019EndtoendOI} proposed an  attention mechanism using a so-called \textit{residual non-local attention block} (RNAB) \cite{zhang2018residual} to capture global dependencies between features. Furthermore, Cheng et al. \cite{cheng2020learned} adopted similar attention-based mechanisms and extended the symbol entropy estimation with a Gaussian mixture model. Their model achieves a performance close to the recent compression standard VVC \cite{jvet2019versatile}.

The intra codec of NN-based video compression is highly related to the image codec. For video compression, Guo has proposed an end-to-end trained codec called \textit{DVC} \cite{lu2019dvc}. The loss function of DVC contains distortion and rate. On top of DVC, Jianping et al. proposed \textit{M-LVC} \cite{lin2020mlvc} with inter-prediction and residual coding parts with increased time-domain dependency utilization. In addition, M-LVC has \textit{MV Refine-Net} and \textit{Residual Refine-Net} to better analyze the information from multi reconstructed frames.  

A particular framework called conditional autoencoder is widely used in image and video compression. A conditional autoencoder can use the auxiliary information to enhance the reconstructed quality. For example, Ladune et al. proposed a novel inter-frame coding method \cite{Ladune2020OpticalFA} that uses two complementary autoencoders, called \textit{MOFNet} and \textit{CodecNet}. MOFNet uses optical flow to predict the next frame. On the other hand, CodecNet is a conditional autoencoder using MOFNet's prediction as an auxiliary information, to better extract the dependencies between original and predicted frame. In the image compression area, Brand et al. propose to use a conditional encoder for image compression in \cite{9244548}. As in that case there is no motion prediction, the conditional encoder uses pixels from the local neighborhood as auxiliary information. The input image is split into $16 \times 16 $ blocks, and an area with a four pixel width around each block is used to condition the encoder.  

In this work, our goal is to design a conditional image autoencoder which can use pixel-wise parallel processing of different color components. Additionally, our conditional autoencoder structure preserves the correlation between the color components.

\label{sec:intro}

\section{Proposed Method}
\begin{figure}[t]
	\centerline{\includegraphics[trim=1.089cm 4.248cm 1.198cm 0cm, clip,width=1.0\linewidth]{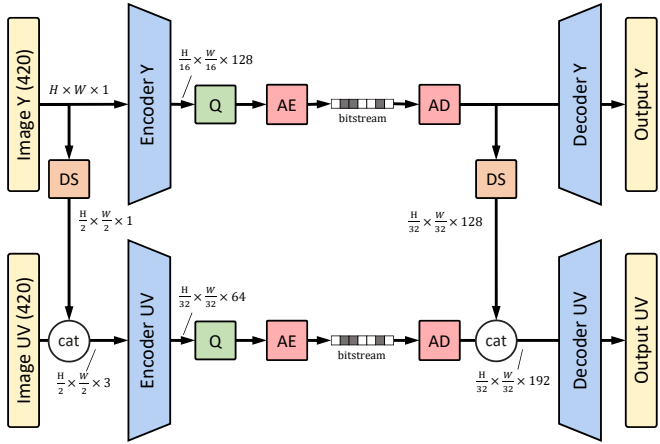}}
	\caption{Example structure of our codec. ``AE'' and ``AD'' indicate arithmetic encoder and arithmetic decoder. ``Q'' denotes the quantizer. ``DS'' stands for down-sampling. ``cat'' is used for channel-wise concatenation. Here, the codec for the Y component uses 128 channels and the one for the UV component uses 64 channels. For simplicity, the hyperprior and the context model are omitted. }
	\label{figfullcodec}
\end{figure}
\subsection{Conditional Color Separation (CCS) }
\label{condres}

The results in the works of \textit{Brand} \cite{9244548} and \cite{Ladune2020OpticalFA} show that conditioning an image autoencoder gives better compression ratio and higher reconstruction quality. However, both approaches have their drawbacks -- in \cite{Ladune2020OpticalFA}, the auxiliary information comes from the entire predicted image, which is not applicable for still image compression. Furthermore, an additional encoder is required to provide the latent presentation of auxiliary information. It increases the model complexity; in \cite{9244548}, the auxiliary information is from the local neighborhood of a block, which hinders parallelized processing. Moreover, the block-based structure is not helpful for the neural network to process the whole image.

For parallelized processing, a straightforward approach is to process the color components of an image independently and in parallel. Unfortunately, such a codec would not be able to exploit the information redundancy between channels. To remedy this, we propose a conditional autoencoder where main and auxiliary information come from different color channels -- i.e., a  \textit{primary color component} is used to assist the processing of a \textit{non-primary color component}. The selection of a primary color component happens beforehand, and then processing of color components happens independently. The additional encoder for providing the auxiliary information in the latent space is not needed in our approach. Thus, our conditional autoencoder can avoid the complexity increase of the additional encoder. In addition, our CCS uses the auxiliary information from the primary color component. It doesn't require any information from the neighborhood of a block. Hence, we can avoid the block-based structure when processing images.

\subsection{Network Architecture} \label{embodiment}
In our CCS, the input data is split into two groups: primary and non-primary components. Each component group has its own dedicated encoder and decoder. In Fig. \ref{figfullcodec} we show an example implementation of CCS for processing an input in YUV420 format. In this format, it is easy to distinguish the primary and non-primary components. Details about the layer structure of each block are given in Table \ref{tab1}.\par

In Fig. \ref{figfullcodec}, the signal size (Height $\times$ Width $\times$ Channel) in each part of the codec is written as ``$H\times W \times C$".  The YUV format has three components -- Y contains the luminance, while U and V contain the chrominance information. In YUV420, U and V components are sub-sampled by a factor of 2 in each direction. Since Y has 4 times more samples, it is selected to be the primary component. Before processing, the non-primary components U and V are concatenated, and the primary component Y is down-sampled to match the height and width of the concatenated UV. Each component is processed by a dedicated  encoder, decoder, hyperprior, auto-regressive context model, arithmetic encoder and arithmetic decoder. Processing of Y happens in parallel with processing of UV. In our example, each convolution layer in the Encoder Y, Decoder Y and Hyperprior Y has 128 channels, while the layers of Encoder UV, Decoder UV and Hyperprior UV have 64. For the purpose of conditioning, the primary component Y is concatenated to UV and fed as an auxiliary information to Encoder UV. Conditioning is also happening in the decoder. There, the latent representation of Y is used as an auxiliary information. The latent Y is down-sampled to match the latent UV, both representations are concatenated and fed together to the Decoder UV.

\begin{table*}[t]
	\centering
	\footnotesize
	\begin{tabular}{ccccccc}
		\toprule
		\textbf{Encoder}&\textbf{Decoder}&\textbf{Hyper Encoder}&\textbf{Hyper Decoder}&\textbf{Context}&\textbf{Gather}\\
		\midrule
		\makecell[t]{RB c$N$ s2 \\RB c$N$ s1\\RB c$N$ s2\\AB c$N$\\RB c$N$ s1\\RB c$N$ s2\\RB c$N$ s1\\Conv: 3×3 c$N$ s2\\AB c$N$ s1}
		
		&\makecell[t]{AB c$M_{Y/UV}$ s1\\RB c$N$ s1\\RBU c$N$ s2\\RB c$N$ s1\\RBU c$N$ s2\\AB c$N$ s1\\RB c$N$ s1\\RBU c$N$ s2\\RB c$N$ s1\\Subpel-Conv: 3×3 c$N$ s2}
		
		&\makecell[t]{Conv: 3×3 c$N$ s1\\Leaky ReLU\\Conv: 3×3 c$N$ s1\\Leaky ReLU\\Conv: 3×3 c$N$ s2\\Leaky ReLU\\Conv: 3×3 c$N$ s1\\Leaky ReLU\\Conv: 3×3 c$N$ s2}
		
		&\makecell[t]{Conv: 3×3 c$N$ s1\\Leaky ReLU\\Subpel-Conv: 3×3 c$N$ s2\\Leaky ReLU\\Conv: 3×3 c$1.5N$ s1\\Leaky ReLU\\Subpel-Conv: 3×3 c$1.5N$ s2\\Leaky ReLU\\Conv: 3×3 c$2N$ s1}
		
		&\makecell[t]{Masked: 5×5 c$2N$ s1}
		
		&\makecell[t]{Conv: 1×1 c$\frac{11}{3}N$ s1\\Leaky ReLU\\Conv: 1×1 c$\frac{10}{3}N$ s1\\Leaky ReLU\\Conv: 1×1 c$3N$ s1}
		\\
		\bottomrule
	\end{tabular}
	\caption{Layer details of our CCS model, $N$ and $M$ indicate the channel number. For Codec Y and Codec UV, $N$ is 128 and 64, respectively. For the first layer of the decoder, $M_Y$ is 128, whereas $M_{UV}$ is set to 192 since the Decoder UV has the concatenated data as input.}
	\label{tab1}
\end{table*}

In Table \ref{tab1}, each row describes one layer of the neural network. ``Conv" prefix means convolutional layer, and the three labels which follow indicate kernel size, channel number, and stride number. The rest of prefixes correspond to the following layer types: ``Subpel-Conv" is a 3x3 sub-pixel convolution for up-sampling; ``Masked" is a mask convolution, as proposed in \cite{oord2016conditional}; ``RB" is a residual block, ``RBU" is a residual block with sub-pixel up-sampling; ``AB" is a self-attention block, as proposed in \cite{cheng2020learned}. In the rest of our text about CCS, we always use the same networks structure, with the only exception of changing the number of the channels. Our implementation is based on Cheng's attention model, proposed in \cite{cheng2020learned}, and we focus our research on frame-work level changes.
\fancyhead[CH]{To be presented at the Picture Coding Symposium (PCS), 7-9 December 2022, San José, California, USA}

\section{Experimental Results}
\subsection{Baseline Model}
Our proposal is concerned with the high-level concept, i.e., the order of the blocks in the framework, rather than the specific implementation of each block. For this reason, we selected a well-known state-of-art codec, the one proposed by Cheng \cite{cheng2020learned} as our baseline. Although this codec is meant for RGB images, our CCS codec works better in YUV420 format. To facilitate comparison to Cheng's original results, we trained two baseline models, the RGB model, hereafter known as \textit{Cheng}, and a model trained on YUV444 data, to which we refer as \textit{Cheng-YUV}. 
\begin{figure}[t]
    \centering
    \includegraphics[width = 1\linewidth]{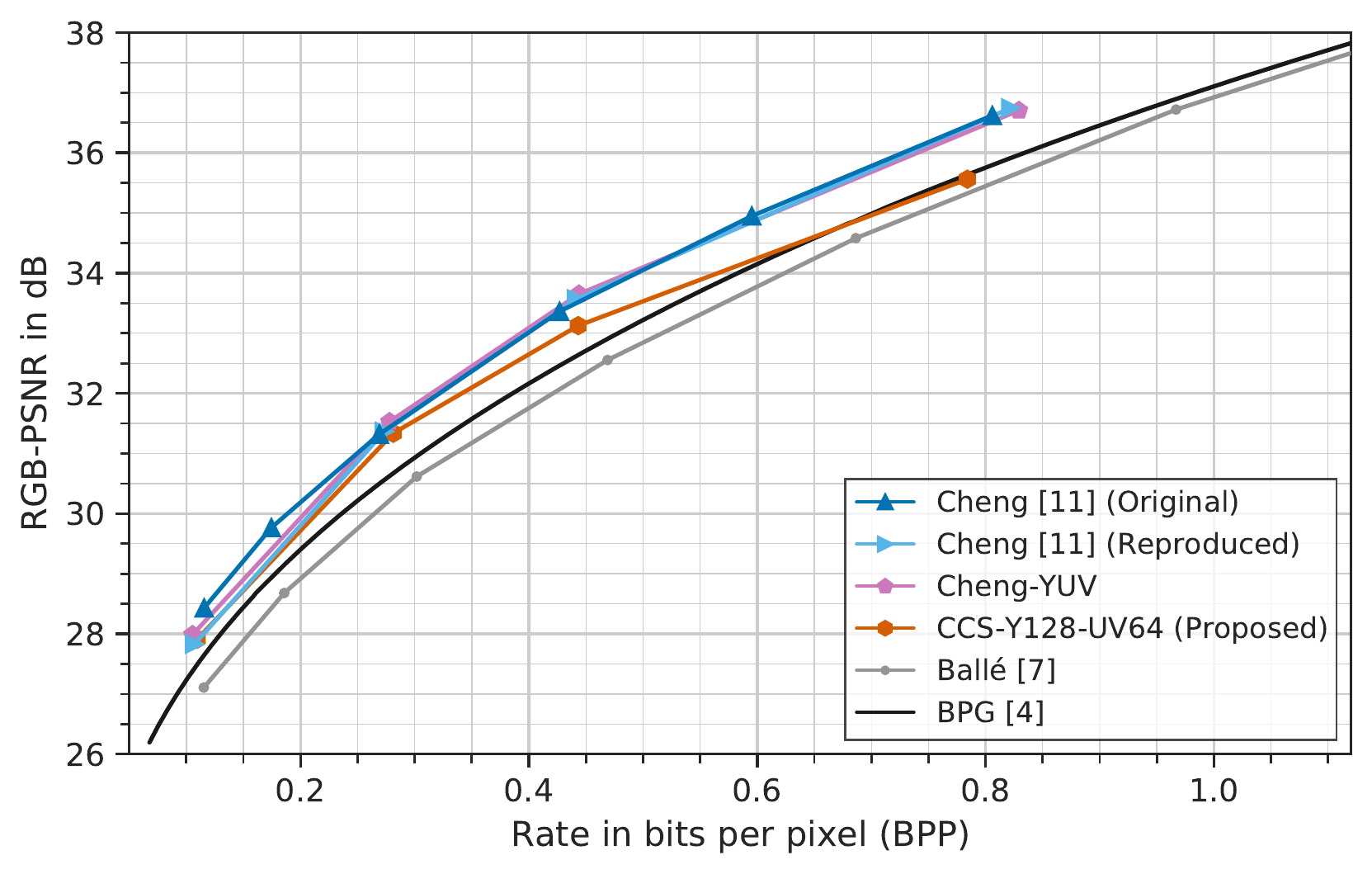}
    \caption{Rate-distortion performance on Kodak dataset for our model and various learning-based and classical codecs. }
    \label{fig:bd_rate}
\end{figure}
\subsection{Conditional Color Separation Model}
\label{ccstest}
In Section \ref{embodiment} we have described an implementation of CCS where the codecs for the primary and non-primary components follow the structure of our baseline model, with the only difference being the number of channels used. We refer to this implementation as \textit{CCS-Y128-UV64}, as it uses 128 channels for the primary Y component and 64 channels for the non-primary UV one. Additionally, we have trained more CCS models using different combinations of channels - \textit{CCS-Y128-UV128} and \textit{CCS-Y64-UV128}. Furthermore, we have trained a model that processes Y and UV separately, i.e., no conditional coding is used between Y and UV. We will use \textit{NC-Y128-UV64} to represent this model. Comparing NC-Y128-UV64 with CCS-Y128-UV64, the performance gain from the condition can be observed.   
\begin{figure}[t]
    \centering
    \includegraphics[width = 1\linewidth]{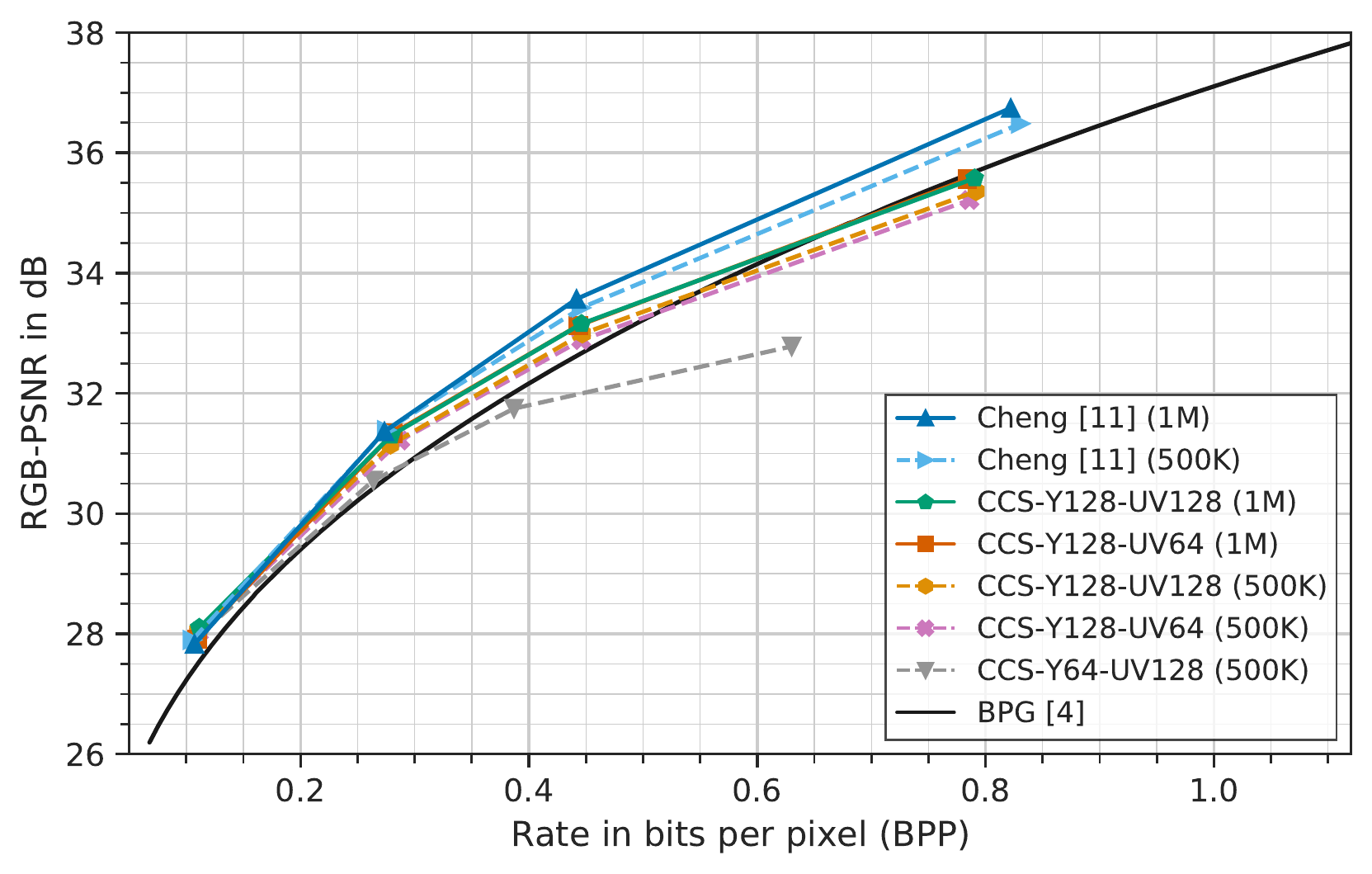}
    \caption{Rate-distortion performance on Kodak dataset for our models with various configurations.}
    \label{fig:placeholder}
\end{figure}
\subsection{Training Strategy}
We selected CompressAI \cite{begaint2020compressai} as our experimental platform for its compact training and evaluation scripts. For our two baseline models (i.e., Cheng and Cheng-YUV), we use quality level of 6, 192 channels, 3 color components and a set of channel number combinations as explained above. For training we use the COCO dataset \cite{coco}. The codec is trained for 1M iterations, and in each iteration we train 16 patches of 256x256 pixels. We trained four checkpoints for each model, each one with different Lagrange multiplier $\lambda$ in the loss function $ L = \lambda \cdot 255^{2} \cdot D + R$, where $D$ is MSE distortion and $R$ is the bit rate. The four checkpoints for each model are trained for $\lambda \in \{0.002,\,0.007,\,0.015,\,0.05\}$.
\begin{table}[t]
\begin{center}
\footnotesize
\begin{tabular}{cc}
\toprule
\textbf{Models}&\textbf{BD-Rate}\\
\midrule
\makecell[c]{BPG \cite{bellard2015bpg}}&\makecell[c]{0.00\%}\\
\makecell[c]{VTM\cite{vtm2019anchor}}&\makecell[c]{-21.08\%}\\
\makecell[c]{Balle et al. \cite{Ball2017EndtoendOI} (autoencoder)}&\makecell[c]{33.91\% }\\
\makecell[c]{Balle et al. \cite{Ball2018VariationalIC} (hyperprior) }&\makecell[c]{10.43\%}\\
\makecell[c]{Balle et al. \cite{Minnen2018JointAA} (hyperprior with context model) }&\makecell[c]{-11.37\%}\\
\makecell[c]{ Cheng et al. \cite{cheng2020learned} (original)}&\makecell[c]{-18.60\%}\\
\makecell[c]{ Cheng et al. \cite{cheng2020learned} (reproduced) 500K}&\makecell[c]{-16.52\%}\\
\makecell[c]{ Cheng et al. \cite{cheng2020learned} (reproduced) 1M}&\makecell[c]{-16.78\%}\\
\makecell[c]{CCS-Y128-UV64 (proposed) 500K}&\makecell[c]{-8.37\%}\\
\makecell[c]{CCS-Y128-UV64 (proposed) 1M}&\makecell[c]{-12.61\%}\\
\makecell[c]{CCS-Y128-UV128 (proposed) 500K}&\makecell[c]{-9.85\%}\\
\makecell[c]{CCS-Y128-UV128 (proposed) 1M}&\makecell[c]{-12.96\%}\\
\makecell[c]{CCS-Y64-UV128 (proposed) 500K}&\makecell[c]{0.82\%}\\
\makecell[c]{NC-Y128-UV64 (proposed) 1M}&\makecell[c]{-10.37\%}\\
\bottomrule
\end{tabular}
\caption{The BD-rate comparison of all our trained models and already published models on Kodak dataset over BPG \cite{bellard2015bpg}. Our models' 500K and 1M iterations training results are shown.} 
\label{Compall}
\end{center}
\end{table}
\begin{table}[t]
\footnotesize
\begin{center}
\scalebox{0.9}{
\begin{tabular}{cccccccc}
\toprule
\textbf{Models}&\textbf{Enc Time}&\textbf{Dec Time}&\textbf{Model Size}&\textbf{KMAC/px}\\
\midrule
\makecell[c]{Cheng \cite{cheng2020learned}}&\makecell[c]{6.46 $s$}&\makecell[c]{11.88 $s$}&\makecell[c]{364.4 $MB$}&\makecell[c]{1027}\\
\makecell[c]{CCS-Y128-UV64}&\makecell[c]{3.66 $s$}&\makecell[c]{9.02 $s$}&\makecell[c]{201.7 $MB$}&\makecell[c]{485}\\
\makecell[c]{CCS-Y128-UV128}&\makecell[c]{4.26 $s$}&\makecell[c]{10.16 $s$}&\makecell[c]{322.0 $MB$}&\makecell[c]{571}\\
\makecell[c]{CCS-Y64-UV128}&\makecell[c]{2.65 $s$}&\makecell[c]{8.23 $s$}&\makecell[c]{199.0 $MB$}&\makecell[c]{229}\\
\makecell[c]{NC-Y128-UV64}&\makecell[c]{3.63 $s$}&\makecell[c]{8.89 $s$}&\makecell[c]{193.0 $MB$}&\makecell[c]{484}\\
\bottomrule
\end{tabular}}
\caption{Encoding and decoding time, and model complexity of our models compared to the model from Cheng et al. \cite{cheng2020learned}.} 
\label{reduTable}
\end{center}
\end{table}
\subsection{Result Evaluation}
The rate-distortion performance of the models trained for 1M iterations on Kodak dataset \cite{kodak} is shown in Fig. \ref{fig:bd_rate}. To make the figure clear, we only show the critical lines in Fig. \ref{fig:bd_rate}. The complete BD-rate comparison of all trained models is shown in Table \ref{Compall}. BD-rate is measured with RGB. The models we trained as baseline (i.e., Cheng and Cheng-YUV) have less than 2\% Bjøntegaard Delta rate (BD-rate) \cite{bjontegaard2001calculation} loss compared to the original Cheng model over BPG \cite{bellard2015bpg}. The mismatch is probably due to our training strategy being different than the one used in \cite{cheng2020learned}. The original model uses different quality models for each $\lambda$, while we use a constant quality of 6 for all $\lambda$ values. Quality 6 indicates a model with 192 channels in every block of the codec. Other differences in our training strategy are -- we used COCO \cite{coco} instead of Vimeo90K \cite{xue2019video} and we used four $\lambda$ values {($\lambda\in \{0.002,\,0.007,\,0.015,\,0.05\}$)} instead of the six originally used ($\lambda\in \{0.0018,\,0.0035,\,0.0067,\,0.0130,\allowbreak\,0.0250,\,0.0483,\,0.0932,\,0.1800\}$).\par

We used the same training strategy to train our CCS models and the reproduced Cheng models. The CCS-Y128-UV64 model has an average 4\% BD-rate loss compared to reproduced Cheng model over BPG \cite{bellard2015bpg}. The reduction of complexity could cause loss, and the BD-rate loss for higher bit rate being larger than the loss for the lower bit rate. However, it is to be noted that the size of the CCS-Y128-UV64 checkpoint is 58\% of the one for Cheng model. Additionally, CCS-Y128-UV64 has 2x faster encoding and 22\% faster decoding speed than the baseline Cheng model.\par

The rate-distortion performance of the 1M and 500K iterations of our trained models on Kodak dataset \cite{kodak} is shown in Fig. \ref{fig:placeholder}. 
As can be seen for training duration of 500K iterations, the increased number of channels in the CCS-Y128-UV128 model over the CCS-Y128-UV64 one leads to 1\% BD-rate gain, while the model size increases by 38\%. If we compare the CCS-Y128-UV128 and CCS-Y64-UV128 models, we see that the smaller number of Y channels leads to 10\% BD-rate loss and 39\% smaller model size. From these results we can conclude that a sufficient number of channels for the primary component helps with the reconstruction quality. On the other hand reducing the channels for the non-primary component affects the quality less, so it is preferred to decrease the channel numbers in this position if a smaller model size is desired. Decreasing the model size additionally helps to achieve shorter processing times and a lower memory consumption. 

Comparing 1M iterations training of CCS-Y128-UV64 and NC-Y128-UV64 model, we observe a 2.24\% BD-rate drop over BPG when not using conditional coding. But the model size of NC-Y128-UV64 is only reduced by 4\%. Thus, the auxiliary information (i.e., condition) for the cross-color components helps us get over 2\% BD-rate gain with a small increasing complexity. For more details we direct the reader to Table \ref{reduTable}.

\section{Conclusion}
We have proposed a learned image compression approach which uses a conditional autoencoder to allow for reduction of memory consumption and computational complexity. Our codec processes primary and non-primary color components in parallel, and uses the primary component as an auxiliary information to help with the encoding and reconstruction of the non-primary component. We refer to this codec as \textit{conditional color separation} (CCS).

Compared to the reproduced autoencoder, which is proposed in \cite{cheng2020learned}, CCS has a 4\% BD-rate loss, however it brings a 42\% model size reduction, 2x faster encoding time and, 22\% faster decoding time. Additionally, our codec allows that data is processed in parallel by multiple GPU cores, which makes it especially suitable for mobile applications.

\bibliographystyle{IEEEbib}
\bibliography{newref}

\end{document}